\newcommand*{\notFOREPJ}{}%
\newcommand*{\FORarXiv}{}%

\ifdefined\FORsinglecolumn
\documentclass[letterpaper,9pt]{scrartcl}
\usepackage{geometry}
\else
\documentclass[%
 aip,
 jmp,%
 amsmath,amssymb,
 reprint,%
]{revtex4-1}
\fi

\usepackage{graphicx}
\usepackage{dcolumn}
\usepackage{bm}
\usepackage{soul}
\usepackage{lineno}
\usepackage{epstopdf}
\usepackage[load-configurations=abbreviations,tight-spacing=true,separate-uncertainty,bracket-numbers = false]{siunitx}
\usepackage[percent]{overpic}
\usepackage{url}

\newcommand{\eV}{\ensuremath{\textrm{eV}}}
\newcommand{\cspeed}{\ensuremath{\mathnormal{c}}}
\newcommand{\dd}{\mathrm d}

\newcommand{\Mimosa}{\ensuremath{\textrm{MIMOSA\,26}}}

\newcommand{\epssut}{\ensuremath{\mathnormal{\varepsilon_{\textrm{SUT}}}}}

\newcommand{\dz}{\ensuremath{\textrm{d}z}}

\newcommand{\dzsut}{\ensuremath{\textrm{d}z_{\textrm{SUT}}}}
\newcommand{\xzero}{\ensuremath{\mathnormal{X}_0}}

\newcommand{\sigmaedge}{\ensuremath{\sigma_{\textrm{edge}}}}

\newcommand{\zsut}{\ensuremath{z_{\textrm{SUT}}}}

\newcommand{\eudet}{\ensuremath{\textrm{EUDET}}}

\setpagewiselinenumbers
\modulolinenumbers[5]

\begin{document}


\ifdefined\FORsinglecolumn
\title{Feasibility of track-based multiple scattering tomography}
\author{Hendrik Jansen,Paul Sch\"utze} 
\else
\title[Track-based Multiple Scattering Tomography]{Feasibility of track-based multiple scattering tomography}

\author{H. Jansen}
 \email{hendrik.jansen@desy.de}
 \affiliation{Deutsches Elektronen-Synchrotron DESY, Notkestr. 85, 22607 Hamburg, Germany}
\author{P. Sch\"utze}
 \email{paul.schuetze@desy.de}
\affiliation{Deutsches Elektronen-Synchrotron DESY, Notkestr. 85, 22607 Hamburg, Germany}

\date{\today}
\fi
\begin{abstract}
We present a tomographic technique making use of a gigaelectronvolt electron beam for the measurement of the spatial material budget distribution of centimetre-sized objects.
With simulation tools originating from high-energy physics applications,
 a test environment replicating a beam telescope is set up to measure the trajectory of electrons traversing a structured aluminium cube with $\SI{6}{\mm}$ edge length. 
The variance of the deflection angle distribution of the electrons undergoing multiple Coulomb scattering at the aluminium cube serves as an estimator for the radon transform of the material budget distribution. 
Basing the sinogram on position-resolved estimators enables the reconstruction of the original object. 
We show the feasibility of the reconstruction of the three-dimensional material budget distribution of the aluminium cube
 by successively imaging two-dimensional distributions of the projected material budget under varying rotation angles. 
Using a filtered back projection, the reconstructed image yields edge resolutions of the order of $\SI{50(2)}{\um}$ and contrasts of about $\num{6.6(2)}$ compared to air, 
 offering a new technique in non-destructive material testing. 
\end{abstract}
\ifdefined\FORsinglecolumn
\else
\fi

\maketitle


\section{Introduction}
\label{sec:intro}
\ifdefined\notFOREPJ
 
Among the many known tomographic techniques, the Computed Tomography (CT)~\cite{Cormack,Hounsfield,Ambrose} allows for the analysis of macroscopic samples of any given material type
 using the transmission/absorption of keV x-rays. 
The transmission of x-rays is governed by the probability to interact with a nucleus in the sample~\cite{HENKE1993181}.
In turn, the transmission probability depends on the ratio of the path length of the photons traversing the sample over the material's characteristic radiation length $\xzero$.
This ratio is also known as the material budget. 
Therefore, the measurand in CT is a function of the radiation length of x-rays, which varies among materials. 
The radiation length is defined as the path length at which a high-energetic electron loses all but a $1/\mathrm{e}$-th fraction of its original energy by emitting bremsstrahlung~\cite{PhysRevD.86.010001}. 
For photons, the mean free path in terms of pair production is $9/7$ of the radiation length, which in turn is a function of the atomic number $Z$ and the density, c.f.\ reference~\cite{ref:PDG-2014}. 
Hence, regions of differing materials or density, i.e.\ differing material budgets, are distinguishable by their transmission probability. 

Available spatial resolutions of CT machines reach down to few hundreds of nanometres at reasonable contrasts~\cite{nanoCT}. 
Image quality suffers for both highly absorbing and highly transmitting samples resulting in low-contrast transmission patterns~\cite{CTcontrastsoft,Cullen:1989wd}. 
Other imaging techniques target the discrimination of functional tissues and pose certain requirements on the sample type itself:
For instance, PET requires biological markers in order to reveal the sample's substructure~\cite{PET},
 MRI is performed on samples containing nuclei with non-zero spin~\cite{MRI}. 
Typical transmission electron microscopy require very thin samples due to the low electron energy~\cite{OATLEY1966181,TEMsimple}. 
However, the technique presented here is not limited to thin samples and does not require non-zero spin materials, much like the CT,
 but is based on the deflection of charged particles rather than on the transmission probability of photons.
 

We propose a tomographic technique based on the measurement of the impact position and scattering angles of charged particles in the GeV-range traversing a sample under test (SUT)
 that allows for the reconstruction of the spatial material budget distribution of the traversed object~\cite{STOLZENBERG2017173}.
The technique is discussed using the example of electrons, but pion-, proton, or muon-based approaches apply equivalently. 
Incoming electrons probe a sample by undergoing multiple Coulomb scattering off the electric field of charged nuclei resulting in an effective angular deflection~\cite{Moliere:1948zz,PhysRev.89.1256} . 
This kink between the incoming and the outgoing electron is accessible by the measurement of the electron's trajectory, or \textit{track}, using pixelated charged-particle sensors in front and behind the sample.
The variance of the angular distribution measured from many probing electrons within a given area is related to the material budget traversed. 
The electron's impact position on the sample is also accessible by its trajectory. 
Position-resolved kink angle distributions for varying rotation angles of the SUT enable the use of an inverse radon transform~\cite{ref:deans2007radon} for image reconstruction. 
We demonstrate the feasibility of the reconstruction of the three-dimensional material budget distribution using the results of a simulation
 of an electron beam in the GeV-range in conjunction with a high-precision particle tracker. 
We therefore term the technique presented in this work a track-based multiple scattering tomography (TBMST). 
 
Similar to the CT, no constraints on the material type are imposed by the TBMST and the amount of the material budget governs the expected signal. 
The underlying physical quantity for the TBMST is the same as for the CT, as the variance of the deflection angle distribution is inversely proportional to the radiation length.
In contrast to the CT, where a fraction of the photons is absorbed, i.e.\ the measurement technique is a position-resolved counting experiment,
 almost all incoming electrons leave the sample volume in the technique proposed herein, therefore the basis of the TBMST is the position measurement of charged particle trajectories. 
In the CT, the image quality for samples larger than a few times its radiation length is limited by the small amount of photons leaving the sample,
 whereas sufficiently energetic electrons completely traverse the sample in a TBMST. 
This might present an advantage of a charged-particle based tomography over a photon-based tomography for sizeable samples of high-Z material,
 which could be exploited in non-destructive material testing. 
The superb position resolution of charged-particle tracking devices also fulfils the need of being able to resolve structures in the micrometre scale~\cite{JansenEPJ}.

\else
 
\fi
%

\section{Simulation set-up}
\label{sec:tscope}
\ifdefined\notFOREPJ
 
For the presented study electron trajectories have been simulated using the AllPix Detector Simulation Framework~\cite{Allpix-github},
 mimicking a so-called beam telescope. 
Such devices usually comprise a set of successively arranged charged-particle sensors in order to measure particle trajectories and are used extensively in charged particle sensor research and development. 
The simulation is set to replicate an electron beam with a momentum of 1.6\,GeV/$\cspeed$ and its physical interactions while traversing matter in terms of scattering
 and energy deposition using the GEANT4 library~\cite{Agostinelli2003250,1610988}.
Appropriately, the simulation includes the energy loss and scattering processes of the propagated electrons in the sensors as well as in the passive materials including the SUT.
Besides the air and the SUT, the simulated set-up contains the beam telescope and simulates their electronic response to a traversing electron. 
The simulation set-up is shown in~Fig.~\ref{fig:datura_sketch} (A).
Depicted are the pixel sensor planes for the precise spatial measurement of the electrons' impact positions, a generic sample under test with its rotation angle $\varphi$,
 the distance $\dz$ between the planes as well as the distance $\dzsut$ between the sample and neighbouring sensor planes. 
The material budget of a homogeneous SUT reads $\epssut(x,y) = l(x,y) / \xzero$, $l$ representing the path length of a beam particle through the SUT at a certain $x$-$y$-position. 
The beam illuminates the entire set-up homogeneously with its particles initially moving parallel to the $z$-axis. 
A configuration of the beam telescope with an equidistant plane-spacing is chosen for this study, cf.~Fig.~\ref{fig:datura_sketch}, with $\dz = \SI{150}{\milli\meter}$ and $\dzsut = \SI{10}{\milli\meter}$.
This article uses a right-handed coordinate system with the $z$-axis along the beam axis and the $y$-axis pointing downwards.
The origin of the coordinate systems is located at the centre of plane\,0. 

\ifdefined\FORarXiv
\begin{figure}[!t]
	\center
	\includegraphics[trim= 50 180 220 30, width=.9\linewidth]{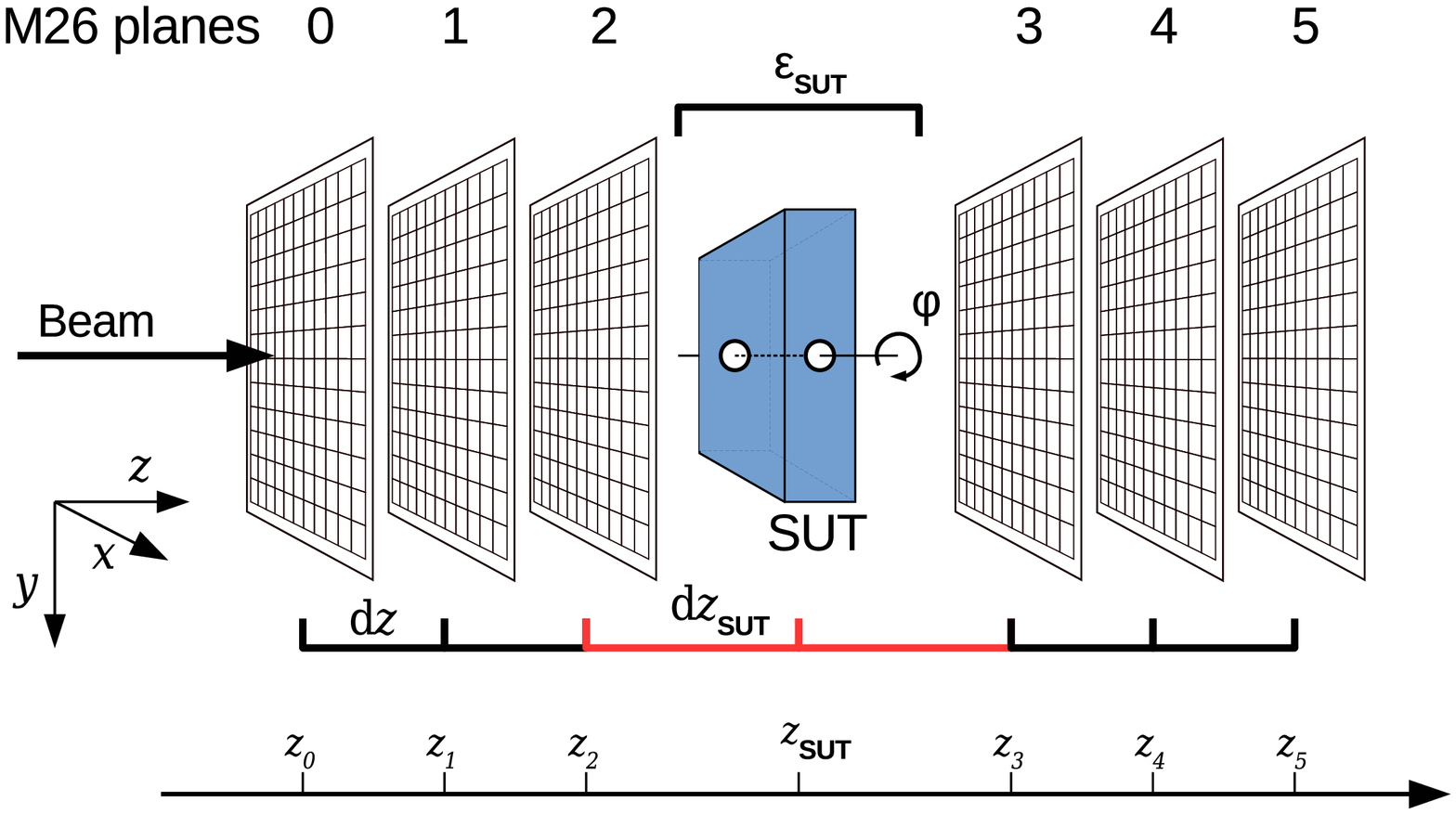}\put(-380,182){(A)}\\
	\includegraphics[trim= 0 0 0 0, width=.45\linewidth]{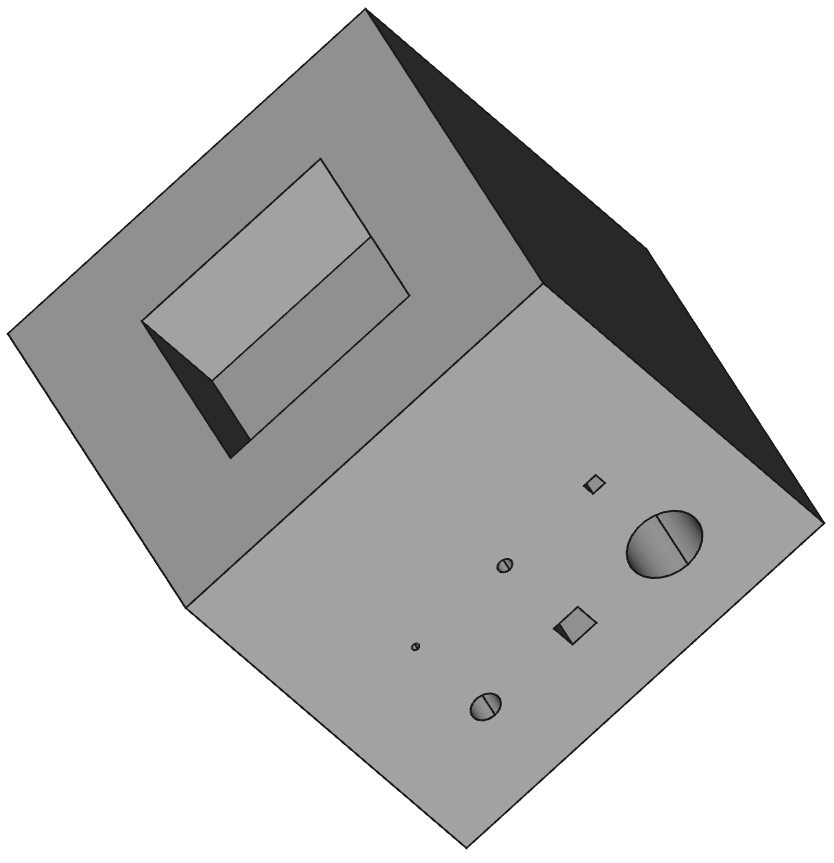}\put(-182,182){(B)}\hspace{3mm}
	\includegraphics[trim= 50 50 50 50, width=.45\linewidth]{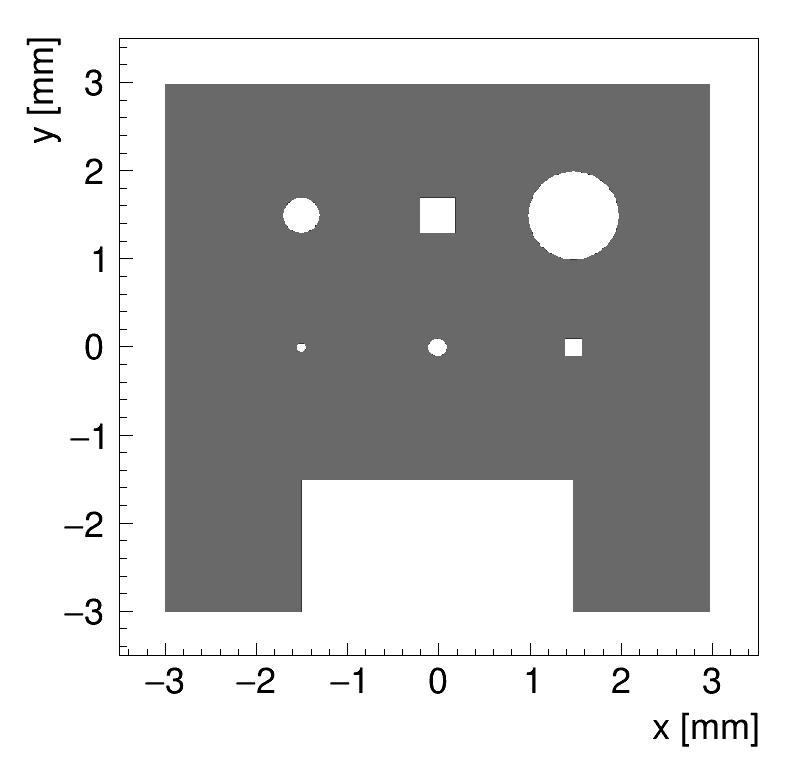}\put(-182,182){(C)}
	\caption[Sketch of the Datura beam telescope]{(A) Sketch of the $\eudet$-type beam telescope with six $\Mimosa$ sensor planes and a generic sample under test in the centre.
	(B) The SUT studied herein is shown in 3D. (C) A cross section through the middle of the SUT.
	}
	\label{fig:datura_sketch}
\end{figure}
\else
\begin{figure}[!t]
	\center
	\ifdefined\notFOREPJ
	\includegraphics[trim= 50 180 220 30, width=.9\linewidth]{figures/sketch_tscope_phantom.eps}\put(-230,120){(A)}\\
	\includegraphics[trim= 0 0 0 0, width=.45\linewidth]{figures/phantomTest.png}\put(-115,105){(B)}\hspace{3mm}
	\includegraphics[trim= 50 50 50 50, width=.45\linewidth]{figures/phantom11.png}\put(-135,105){(C)}
	\else
	\includegraphics[trim= 50 180 220 30, width=.9\linewidth]{sketch_tscope_phantom.eps}\put(-230,120){(A)}\\
	\includegraphics[trim= 0 0 0 0, width=.45\linewidth]{phantomTest.png}\put(-115,105){(B)}\hspace{3mm}
	\includegraphics[trim= 50 50 50 50, width=.45\linewidth]{phantom11.png}\put(-135,105){(C)}
	\fi
	\caption[Sketch of the Datura beam telescope]{(A) Sketch of the $\eudet$-type beam telescope with six $\Mimosa$ sensor planes and a generic sample under test in the centre.
	(B) The SUT studied herein is shown in 3D. (C) A cross section through the middle of the SUT.
	}
	\label{fig:datura_sketch}
\end{figure}
\fi

The SUT consists of a structured aluminium block allowing for the investigation of the potential and limits of this track-based multiple scattering tomography. 
Figures~\ref{fig:datura_sketch} (B) and (C) show a 3D illustration and a cross-section of the SUT at a rotation angle $\varphi = \SI{90}{\degree}$, respectively. 
The SUT is a cube with an edge length of $\SI{6}{\milli\meter}$ and a rectangular cut-out of $\SI{3}{\milli\meter}\times\SI{3}{\milli\meter}\times\SI{1.5}{\milli\meter}$ at the bottom side. 
This enables the measurement of the contrast within a defined region of the material. 
Additionally, the sample features squared and round holes of $\SI{0.1}{\milli\meter}$ to $\SI{1}{\milli\meter}$ in size and diameter allowing for the testing of minimal resolvable feature sizes. 

The beam telescope used in the simulation represents a $\eudet$-type beam telescope, with two of these devices available at the DESY Test Beam Facility~\cite{DESYtb},
 and is described in detail in reference~\cite{JansenEPJ}.
It comprises six pixel detector planes equipped with fine-pitch $\Mimosa$ sensors~\cite{HuGuo2010480}.
Each $\Mimosa$ sensor consists of $\SI{18.4}{\um}\,\times\,\SI{18.4}{\um}$ sized pixels arranged in 1152 columns and 576 rows,
thus covering $\SI{21.2}{\mm}\,\times\,\SI{10.6}{\mm}$ in total.
The detector response is simulated by forming patterns of one or more registered pixels using the impact position of the simulated electron trajectories on the sensors
 based on their measured response~\cite{1748-0221-11-12-C12031}.
The average intrinsic resolution, i.e.\ the average spatial accuracy with which the impact position of the traversing electron can be reconstructed at a sensor,
 has been measured to be $\SI{3.24}{\um}$~\cite{JansenEPJ}. 
With a sensor thickness of $\SI{50}{\um}$ protected by two Kapton foils of the same thickness,
 the material budget of the beam telescope planes are kept as low as possible in order to achieve an excellent track resolution even at beam momenta of a few GeV/$\cspeed$.
The track resolution is defined as the spatial resolution of the electrons track at a given point along its trajectory. 
The track resolution of $\SI{2.0(1)}{\um}$ for a 6\,mm aluminium cube is considerably smaller than the edge resolution, which is discussed in section~\ref{sec:discussion}. 

For all material, electrons are simulated to undergo multiple Coulomb scattering causing an effective angular deflection.
This distribution of scattering angles is centred around zero
 and its variance depends on the electron energy and the radiation length of the matter traversed~\cite{Moliere:1948zz,PhysRev.89.1256,ref:scatteringhighland,Berger}.
The variance of the angular distribution $\Theta_{0}^{2}$ predicted by Highland’s approximation of the Molière theory for single scatterers evaluates to

\begin{equation}
\label{eq:multiplescattering_a}
\Theta_{0}^{2} = \left( \frac{\SI{13.6}{\mega\eV}}{\beta \cspeed p} \cdot z \right)^2
\cdot \varepsilon
\cdot \big( 1 + 0.038 \cdot \ln{\left( \varepsilon \right) } \big)^2 \,,
\end{equation}

\noindent 
with $\beta c$, $p$ and $z$ representing the velocity, the momentum and the charge number of the traversing electron. 
The material budget $\varepsilon$ is defined as the ratio of the path length inside the material and the material's radiation length $\varepsilon = l / \xzero$. 
An accuracy of 11\,\% for all atomic numbers $Z$ is reported for the Gaussian approximation of the inner 98\,\% of the angular distribution
 for scatterers of $0.001 \leq \varepsilon \leq 100$~\cite{ref:scatteringhighland}. 




The simulation is produces sets of trajectories that are subdivided into \textit{events}. 
We define an event as one cycle of simulation with a small number of traversing electrons, analogous to the data produced in experiments using $\eudet$-type beam telescopes at the DESY Test Beam Facility. 
The data retrieved from the simulation~\cite{schuetze_data_TBMST}, covering a total of $\num{280e6}$ simulated electrons, consists of a list of registered sensor pixels per such an event.
A pixel is registered in case a beam particle is simulated to pass through it, and has a certain probability to be registered if the neighbouring pixel is traversed. 

\else
 
\fi

\section{Data preparation and image reconstruction}
\label{sec:datatomo}
\ifdefined\notFOREPJ
 
In order to be used for image reconstruction, the simulated data are processed as follows.
Registered pixels that adjoin are combined to form a so-called cluster. 
Due to the 
binary readout of the $\Mimosa$ sensors, which is reflected in the simulation, a simple geometrical interpolation of the cluster centre is performed, which is defined as the reconstructed hit position.
The hits are translated from two-dimensional entities on the individual beam telescope planes into three-dimensional entities in the global frame of reference.
A single beam particle usually produces six clusters, and hence hits, one per traversed sensor plane.

In order to find hits in the six planes originating from the same beam particle
 so-called \textit{triplets} are built based on hits in the first three planes (upstream) and the last three planes (downstream) separately, as depicted in Fig.~\ref{fig:sketch_TRIPLET}: 
First, doublets are defined by straight lines from all hits in plane 0 to all hits in plane 2. 
A valid triplet is found, if a reconstructed hit in plane\,1 around the doublet interpolation to plane\,1 is present within a distance smaller than $d_{\textrm{val}}$. 
Triplet isolation is ensured by rejecting all triplets whose extrapolation to $\zsut$ approaches other triplet extrapolations of the same event by a distance smaller than $d_{\textrm{iso}}$:
if any triplet is too close to another, both are being discarded ensuring a clean dataset. 
This procedure is repeated accordingly for the downstream telescope planes. 
Finally, six-tuples are formed by matching isolated triplets from the up- and downstream telescope planes. 
A valid six-tuple is defined by a pair of triplets if they intersect within a radius of $d_{\textrm{match}}$ at the SUT's transversal plane. 
In a simplistic model, two straight lines, one originating from the upstream triplet and one from the downstream triplet, describe the track of the beam particle,
 with a single kink allowed at the SUT. 
From this simple model the effective kink angle caused by the multiple Coulomb scattering is extracted as the angle between the upstream and downstream triplets of valid six-tuples.

\begin{figure}[!t]
	\center
	\ifdefined\notFOREPJ
	\includegraphics[trim= 50 150 190 20, width=.8\linewidth]{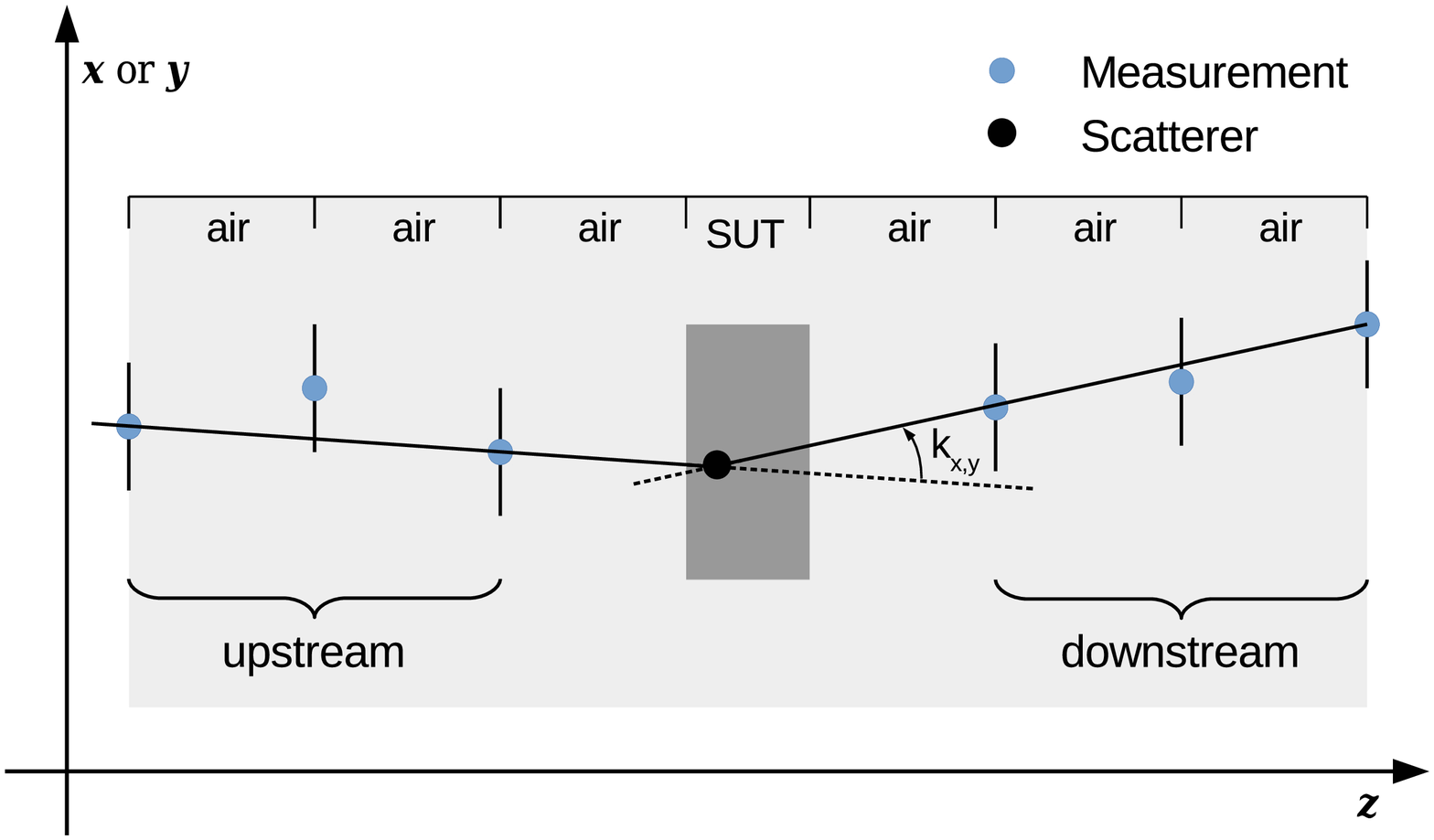}
	\else
	\includegraphics[trim= 50 150 190 20, width=.8\linewidth]{sketch_TRIPLET.eps}
	\fi
	\caption[Sketch of the GBL track]{Sketch of the triplet method.
	Triplets are constructed for both the upstream and downstream sensor planes, their difference in slope yielding an estimate of the effective scattering angle at the SUT.
	}
	\label{fig:sketch_TRIPLET}
\end{figure}


Accumulating many six-tuples, the variance of the kink angle distribution encodes the information about the material budget in the SUT. 
In fact, the solid angle can be decomposed into two projections along perpendicular dimensions and hence a two-dimensional measurement is performed.
Appropriately, we chose the axes parallel to the sensor geometry, i.e.\ along the $x$- and $y$-direction.
Due to the quantum mechanic nature of the scattering process, the kink angles $k_x$ and $k_y$ are expected to be uncorrelated for a single measurement and a single particle.
However, the variance of the two distributions for a given SUT area are expected to fully correlate within uncertainties,
 which in principle enables the calculation of two independent estimates of the material budget distribution.
 
With the method described above, two-dimensional images are acquired that represent the position-resolved variance of the scattering angle distribution,
 and therefore an estimator for the material budget projected onto the $x$-$y$-plane. 
Subsequently, these images are split into vertical slices. 
The horizontal size of a slice is independent of the sensors' pixel size and defines the width (size in $x$) of the so-called voxels,
 the cells of the three-dimensional distribution to be reconstructed. 
The vertical cell size chosen defines the voxel's depth and height (sizes in $y$ and $z$).

A sinogram is a collection of multiple one-dimensional projections of a two-dimensional density distribution, which is a binned representation of the radon transform of the original distribution \cite{ref:RadonOrig}.
In order to fully reconstruct the sample's material budget distribution,
 the simulation and analysis are repeated for different rotation angles of the sample and the corresponding vertical slices from all angles are combined to form such a sinogram.
In our case, each data point in the sinogram is given by the variance of a multiple scattering angle distribution and is therefore, according to eq.~(\ref{eq:multiplescattering_a}),
 not linearly dependent on the material budget.
It is therefore not a perfect estimator for the radon transform of the material budget distribution
 
\begin{equation}
\label{eq:thetaRadon}
\Theta^2(L) =  \left( \frac{\SI{13.6}{\mega\eV}}{\beta \cspeed p} \cdot z \right)^2 \int_{L} \frac1{X_0(x,y,z)}\,|\dd s|,
\end{equation} 

\noindent
but serves as a sufficiently accurate one to prove the concept. 
Therefore, considering a certain slice of the SUT, an inverse radon transform of the sinogram yields a reconstruction of the two-dimensional material budget distribution
 and the combination of reconstructions from adjacent slices results in a three-dimensional image~\cite{ref:deans2007radon}.
Whereas the forward radon transform for a known density distribution is well defined, the inverse radon transform can only be approximated.
For this work, the open source software package scikit-image~\cite{ref:scikitWebpage,ref:scikitArticle} is used,
 which is capable of performing a filtered back projection based on the central-slice theorem~\cite{ref:deans2007radon}.
 

\else
 
\fi

\section{Results}
\label{sec:results}
\ifdefined\notFOREPJ
 
\ifdefined\FORarXiv
\begin{figure}[!tb]
	\center
	\includegraphics[trim= 0 0 0 0, width=.48\linewidth]{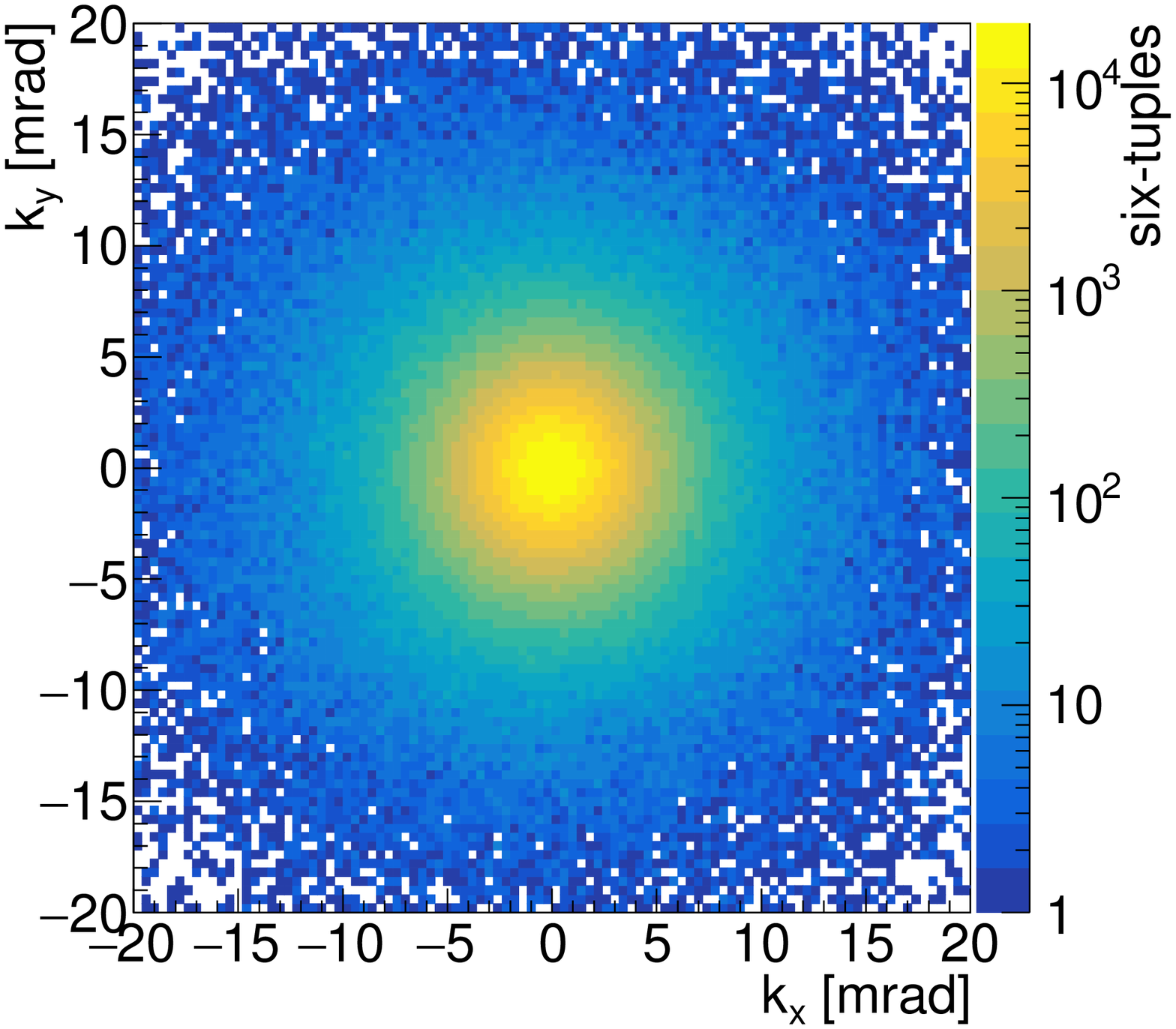}\put(-182,175){(A)}
	\hspace{0.02\linewidth}
	\includegraphics[trim= 0 0 0 0, width=.48\linewidth]{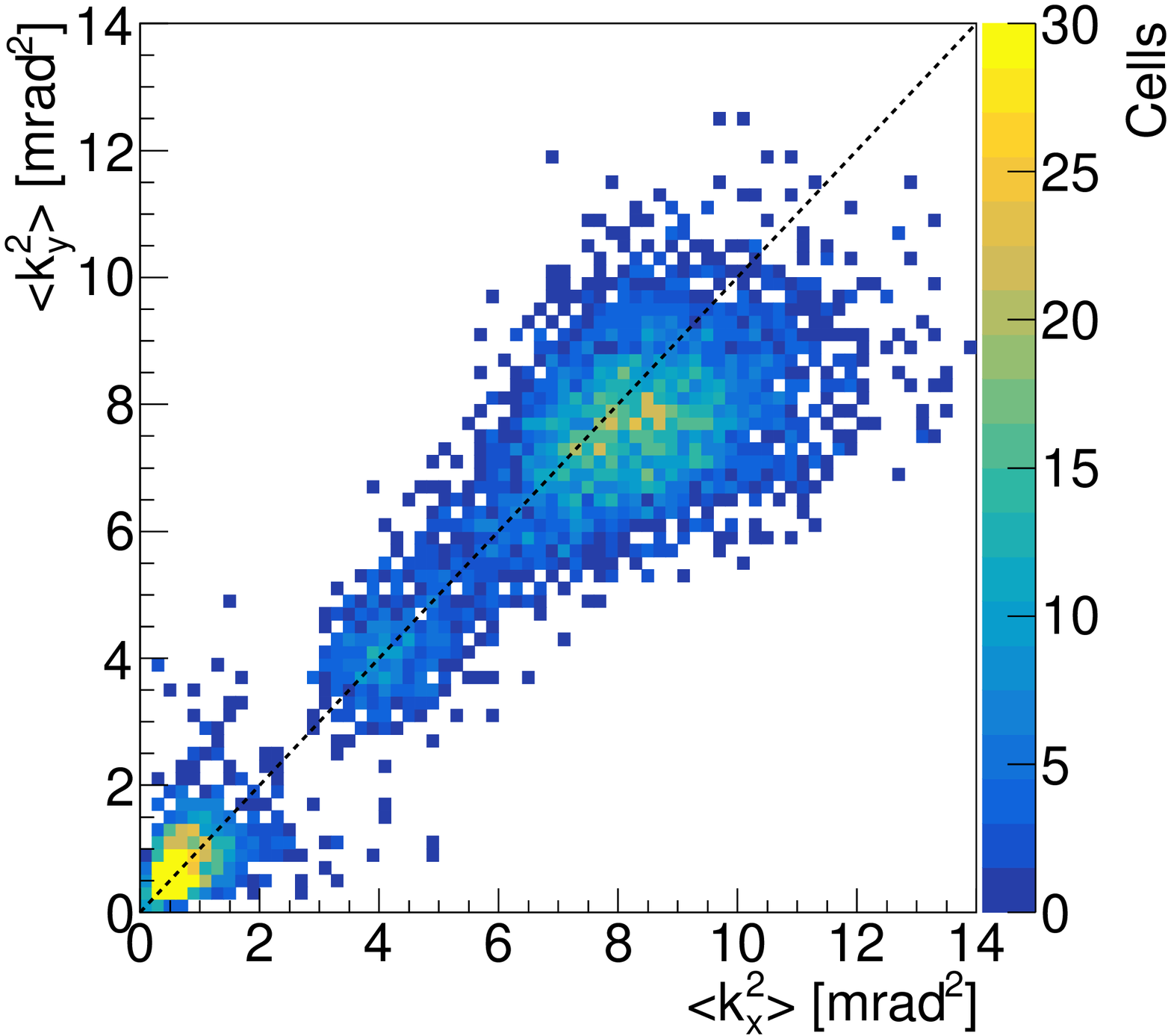}\put(-182,175){(B)}
	\caption[kx ky]{
	(A) A scatter plot of the measured multiple scattering angle per track in the $x$- and $y$-direction. 
	(B) The variances in the $x$- and $y$-direction for single cells of the 2D scattering angle distributions are plotted against each other.
	}
	\label{fig:correlations}
\end{figure}
\else
\begin{figure}[!tb]
	\center
	\ifdefined\notFOREPJ
	\includegraphics[trim= 0 0 0 0, width=.48\linewidth]{figures/kinkCorrelation.eps}\put(-110,108){(A)}
	\hspace{0.02\linewidth}
	\includegraphics[trim= 0 0 0 0, width=.48\linewidth]{figures/meanKinkCorrelation.eps}\put(-110,108){(B)}
	\else
	\includegraphics[trim= 0 0 0 0, width=.48\linewidth]{kinkCorrelation.eps}\put(-110,108){(A)}
	\hspace{0.02\linewidth}
	\includegraphics[trim= 0 0 0 0, width=.48\linewidth]{meanKinkCorrelation.eps}\put(-110,108){(B)}
	\fi
	\caption[kx ky]{
	(A) A scatter plot of the measured multiple scattering angle per track in the $x$- and $y$-direction. 
	(B) The variances for single cells of the 2D scattering angle distributions are plotted against each other.
	}
	\label{fig:correlations}
\end{figure}
\fi

During data preparation, the  following parameters have been used in order to find, isolate and match upstream to downstream triplets:
$d_{\textrm{val}} = \SI{100}{\um}$,
$d_{\textrm{iso}} = \SI{70}{\um}$,
and $d_{\textrm{match}} = \SI{35}{\um}$.
The resulting kink angles $k_x$ and $k_y$ per track at the SUT are shown in a scatter plot in Fig.~\ref{fig:correlations}~(A) for a given rotation angle at $\varphi = \SI{90}{\degree}$ 
 and show no correlation, as expected.
At the same time, the widths of the scattering angle distribution for the $x$- and $y$-direction in single cells of the two-dimensional projection correlate with a Pearson correlation factor of 0.94,
 cf.~Fig.~\ref{fig:correlations}~(B).
Cells containing only air group at the lower left corner, cells containing 6\,mm of aluminium at the upper right. 

There are at least to possibilities to calculate an estimate for $\Theta_{0}^{2}$: 
 (1) The rms of the kink angle distribution is evaluated and squared. 
 (2) The mean of the squared kink angle distribution is evaluated. 
The latter is valid, as the kink angle distribution is symmetric around zero and resembles a normal distribution. 
Electrons traversing the upstream sensors near to the sensor edges might be scattered out of the acceptance area of the downstream sensor planes at the SUT causing a geometrical cut-off of the kink angle distribution.
Therefore, the usage of the mean squared kink angle shows a more uniform performance in comparison to a direct evaluation of the rms of the angular distributions. 
In order to increase statistics, both measurements from the $x$- and the $y$-direction, $k_x$ and $k_y$, are combined into $<k_x^2 + k_y^2>$,
 which is mathematically equivalent to averaging the means of the two squared angles: $<k_x^2 + k_y^2>/\,2 = \,<<k_x^2> + <k_y^2>>$.

\ifdefined\FORarXiv
\begin{figure}[!tb]
	\center
	\includegraphics[trim= 0 0 0 0, width=.48\linewidth]{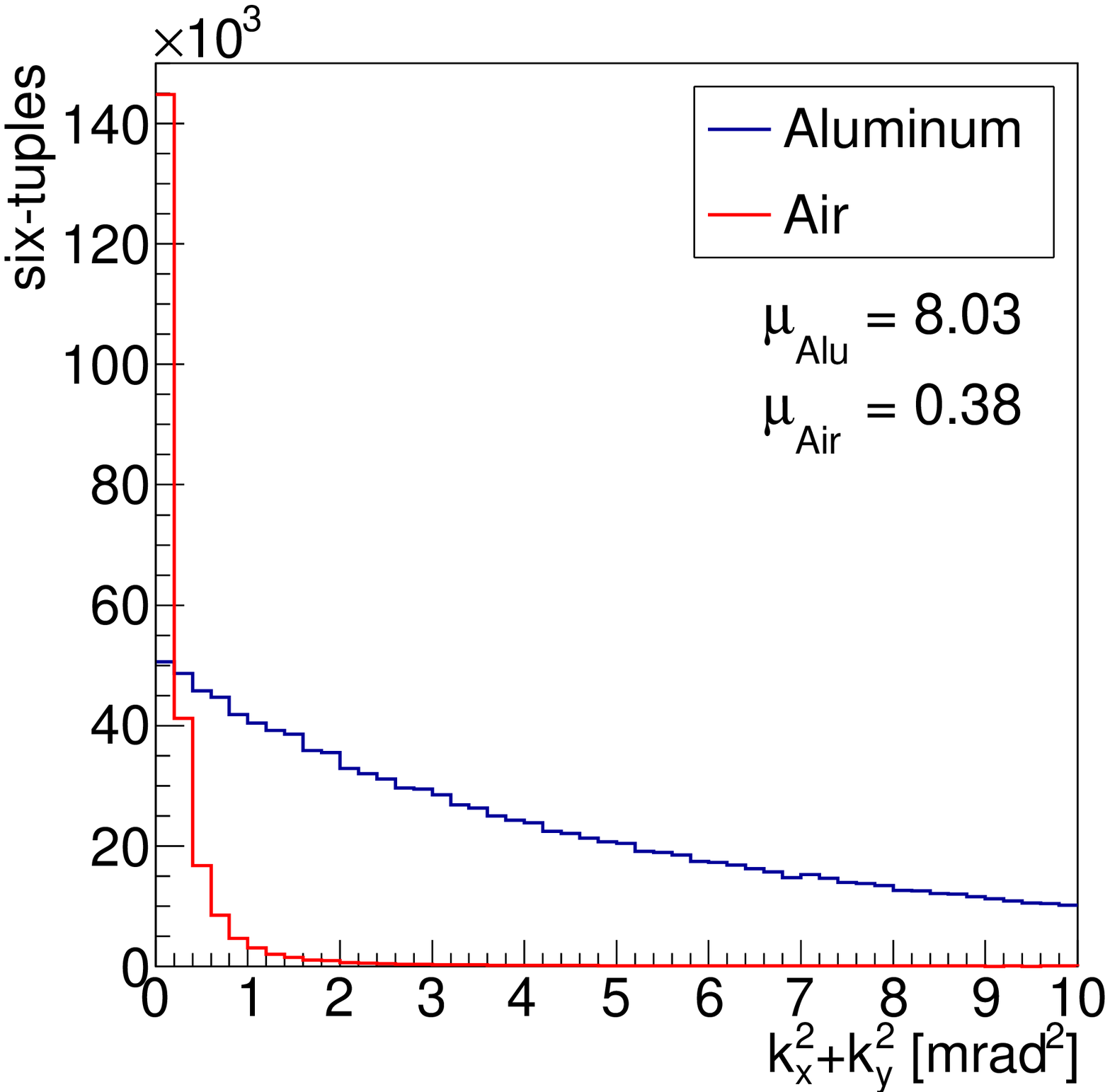}\put(-182,175){(A)}
	\hspace{0.02\linewidth}
	\includegraphics[trim= 0 0 0 0, width=.48\linewidth]{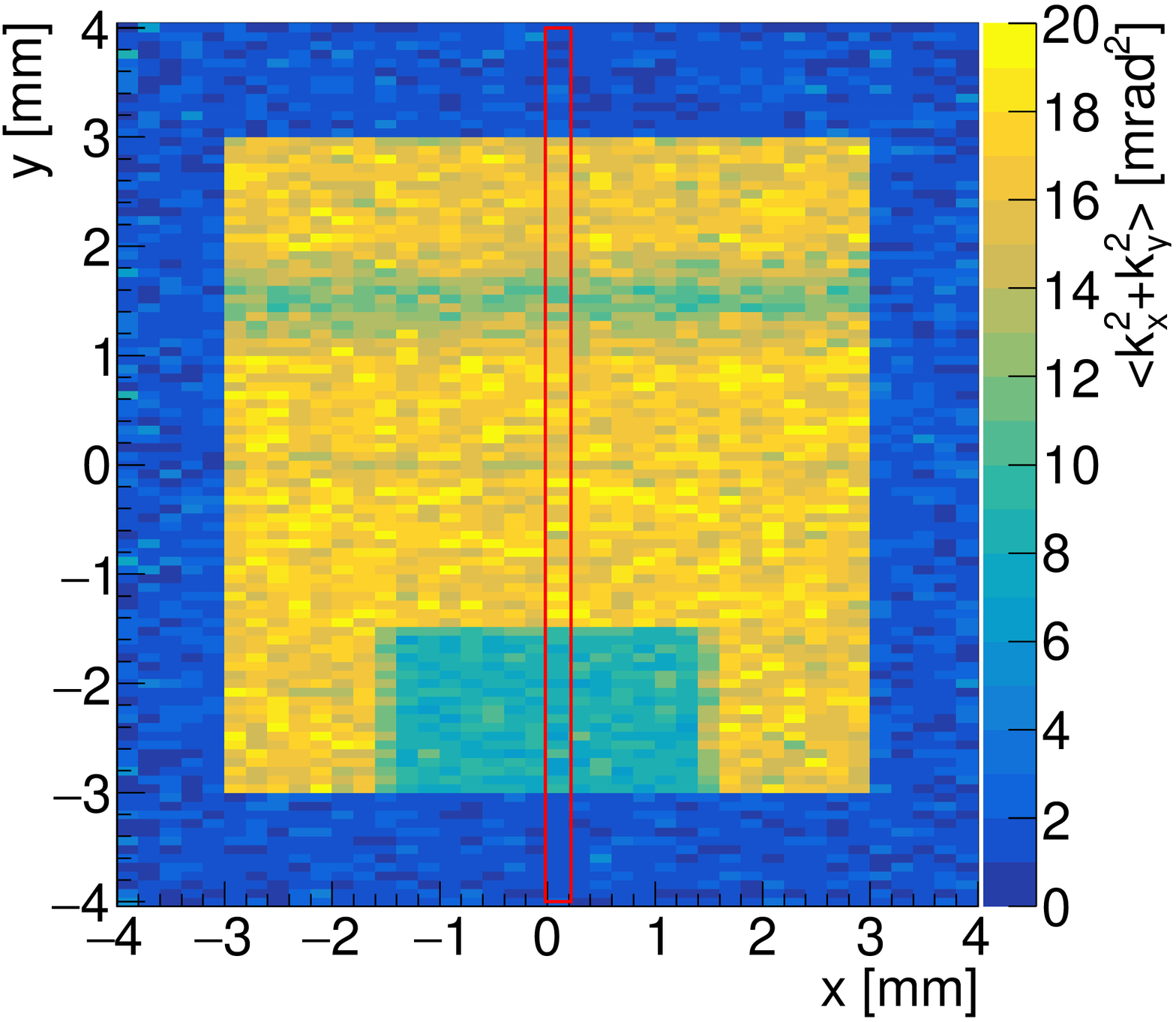}\put(-182,175){(B)}\\
	\includegraphics[trim= 0 0 0 0, width=.48\linewidth]{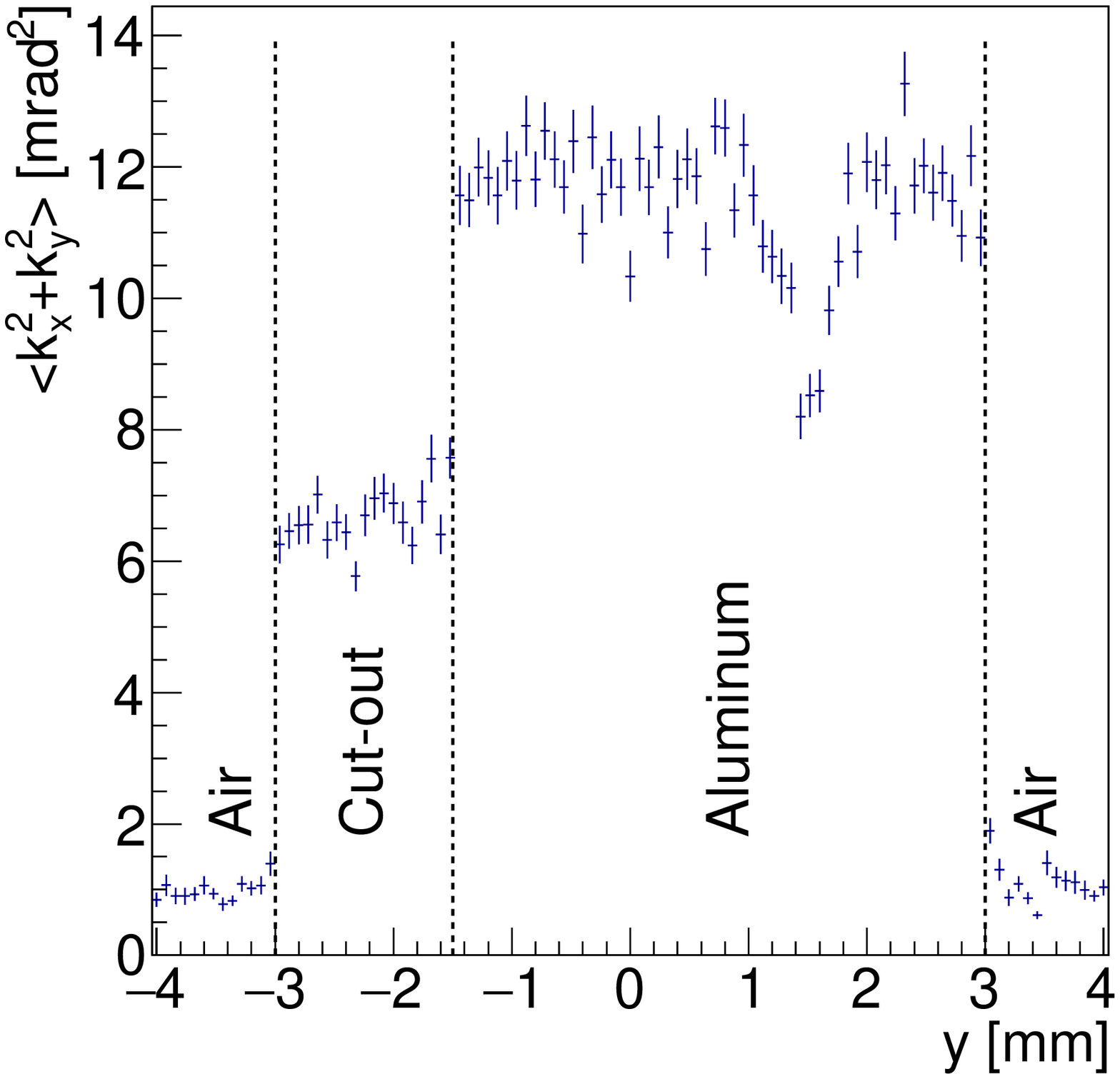}\put(-182,175){(C)}
	\hspace{0.02\linewidth}
	\includegraphics[trim= 0 0 0 0, width=.48\linewidth]{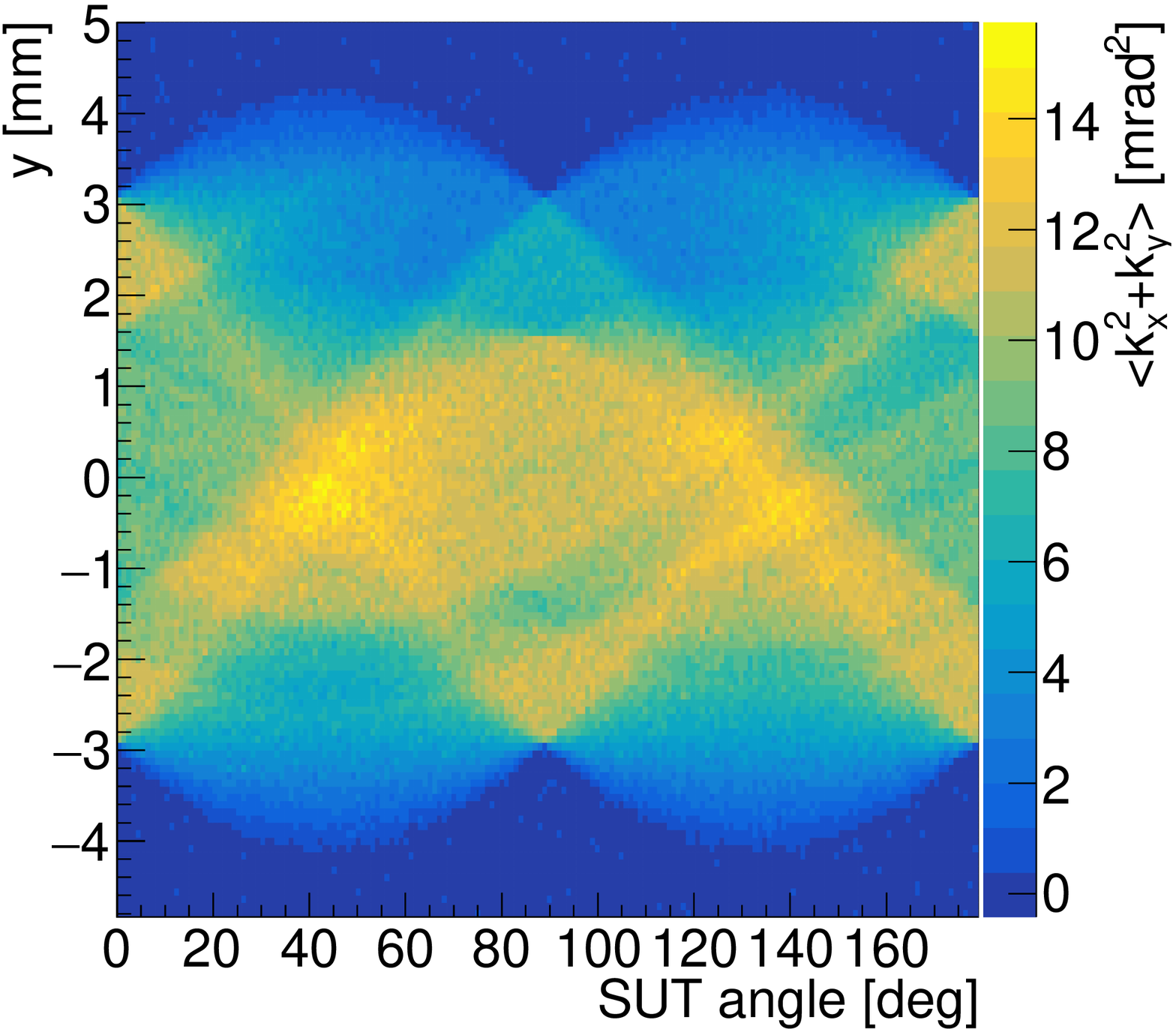}\put(-182,175){(D)}
	\caption[material image and slice]{(A) The squared kink angle is shown for air and aluminium.
	(B) A two-dimensional projection of the material budget in terms of the mean of the squared kink angle distributions. 
	(C) The material budget taken from the red box in (B) is shown along the $y$-direction. 
	(D) The sinogram shows the material budget from (C) concatenated for various rotation angles, i.e.\ in the $x$-$\varphi$-plane.
	}
	\label{fig:2D}
\end{figure}
\else
\begin{figure}[!tb]
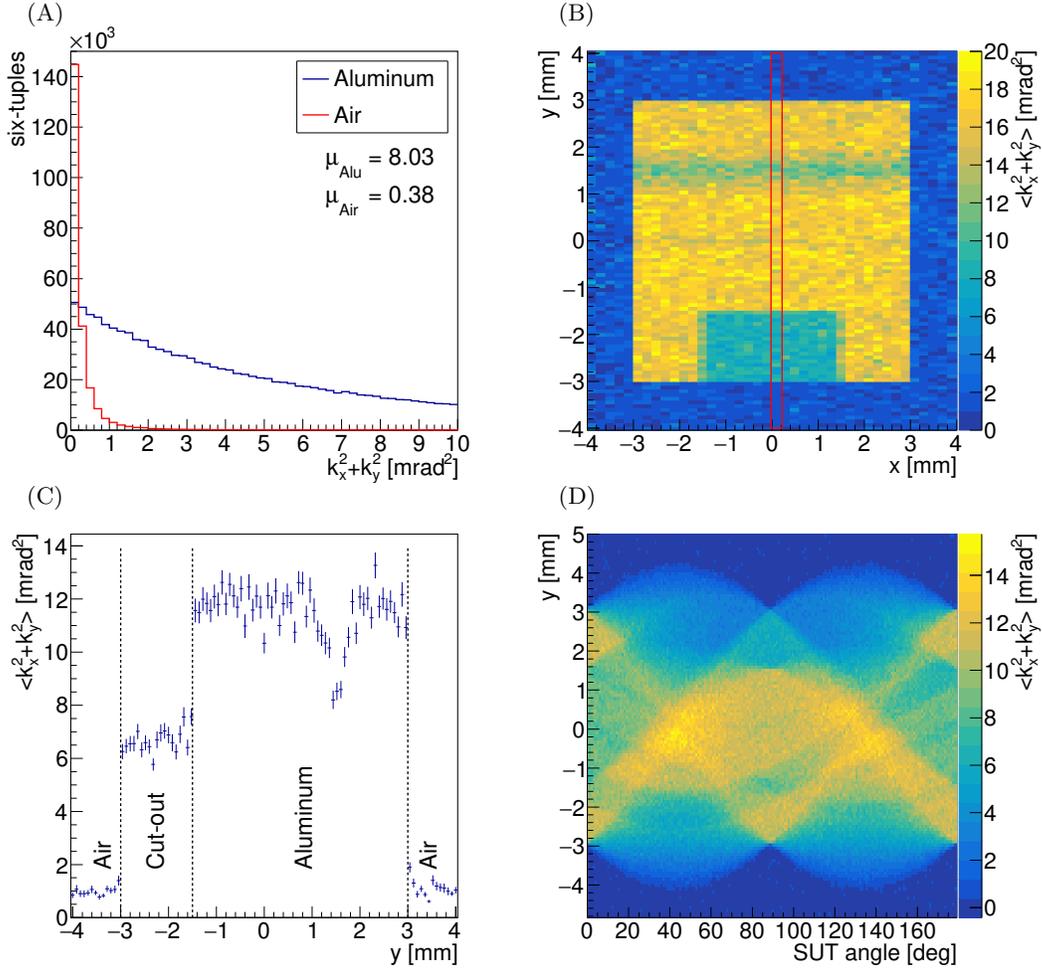

	\center
	\ifdefined\notFOREPJ
	\includegraphics[trim= 0 0 0 0, width=.48\linewidth]{figures/kx2.eps}\put(-110,110){(A)}
	\hspace{0.02\linewidth}
	\includegraphics[trim= 0 0 0 0, width=.48\linewidth]{figures/2Dxy.eps}\put(-110,110){(B)}\\
	\includegraphics[trim= 0 0 0 0, width=.48\linewidth]{figures/slice.eps}\put(-110,108){(C)}
	\hspace{0.02\linewidth}
	\includegraphics[trim= 0 0 0 0, width=.48\linewidth]{figures/sinogram.eps}\put(-110,108){(D)}
	\else
	\includegraphics[trim= 0 0 0 0, width=.48\linewidth]{kx2.eps}\put(-110,110){(A)}
	\hspace{0.02\linewidth}
	\includegraphics[trim= 0 0 0 0, width=.48\linewidth]{2Dxy.eps}\put(-110,110){(B)}\\
	\includegraphics[trim= 0 0 0 0, width=.48\linewidth]{slice.eps}\put(-110,108){(C)}
	\hspace{0.02\linewidth}
	\includegraphics[trim= 0 0 0 0, width=.48\linewidth]{sinogram.eps}\put(-110,108){(D)}
	\fi
	\caption[material image and slice]{(A) The squared kink angle is shown for air and aluminium.
	(B) A two-dimensional projection of the material budget in terms of the mean of the squared kink angle distributions. 
	(C) The material budget taken from the red box in (B) is shown along the $y$-direction. 
	(D) The sinogram shows the material budget from (C) concatenated for various rotation angles, i.e.\ in the $x$-$\varphi$-plane.
	}
	\label{fig:2D}
\end{figure}
\fi

According to eq.~(\ref{eq:thetaRadon}) the mean squared kink angle yields an estimate of the material budget. 
Figure~\ref{fig:2D}~(A) clearly shows the difference between the angular distributions for scattering in air and in $\SI{6}{\milli\meter}$ of aluminium.
An image of the mean squared kink angles in the $x$-$y$-plane for a fixed rotation angle of $\varphi = \SI{90}{\degree}$ is shown in Fig.~\ref{fig:2D}~(B) 
 with a cell size of $200\times\SI{80}{\um^2}$.
The SUT clearly protrudes from the surrounding air, the $\SI{3}{\mm}\times\SI{1.5}{\mm}$ cut-out at the bottom side and the larger holes around $y=\SI{1.5}{\mm}$ are visible. 
Rather small deviations from the solid aluminium are visible around the location of the smaller holes at $y=\SI{0}{\mm}$.

Such material images are constructed for every rotational angle of the SUT. 
For each step, a slice is cut out, exemplified at $0 \leq x < \SI{200}{\um}$ as a red box in Fig.~\ref{fig:2D}~(B). With the slices having a vertical cell size of $\SI{80}{\um}$,
 the voxel size is therefore defined to be $200\times80\times\SI{80}{\um^3}$.
The material budget along the slice in the $y$-direction is shown in Fig.~\ref{fig:2D}~(C). 
A high signal-to-noise ratio is readily visible comparing the regions of air with those of aluminium. 
The projection reveals sharp edges at the borders at $x = \pm \SI{3}{\mm}$ as well as a clear transition from the cut-out to the solid region at $y = -\SI{1.5}{\mm}$.

Slices of the material image for 180 rotational angles of the SUT from 0 to 179 degrees form the sinogram presented in Fig.~\ref{fig:2D}~(D).
This sinogram serves as input for an inverse radon transform, with the result shown in Fig.~\ref{fig:reco},
 representing the reconstructed material budget distribution in the $y$-$z$-plane.
Readily visible are the holes of the upper row, as well as the right and the middle hole of the lower row. 
Merely the smallest hole of $\SI{0.1}{\milli\meter}$ diameter is not visible at the chosen slice width of 0.2\,mm. 
Note that the holes in the lower row neither showed a significant signal in the $x$-$y$- nor in the sliced $y$-projection in Fig.~\ref{fig:2D}~(B) and (C),
 but can be recovered by combining slices from all rotational angles.

\begin{figure}[!tb]
	\center
	\ifdefined\notFOREPJ
	\includegraphics[trim= 0 0 0 0, width=.98\linewidth]{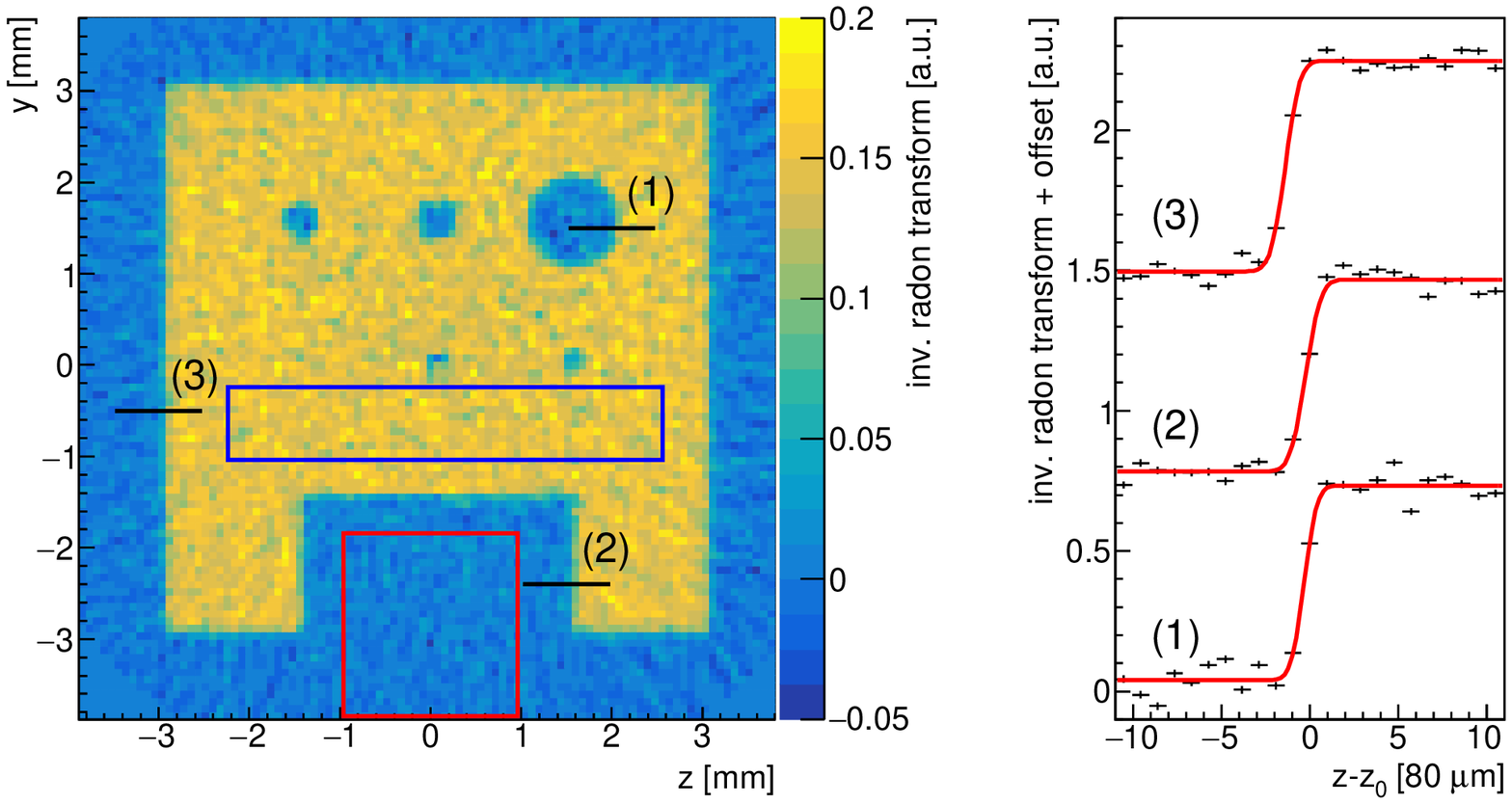}
	\else
	\includegraphics[trim= 0 0 0 0, width=.98\linewidth]{edges.eps}
	\fi
	\caption[reco]{(left) An inverse radon transform on the sinogram (Fig.~\ref{fig:2D}~(D)) yields the reconstructed material distribution in the $y$-$z$-plane. 
	The coloured boxes indicate the regions used for the determination of the contrast, blue for aluminium, red for air. 
	The bracketed numbers reference horizontal lines at which the edge resolution is evaluated, shown on the right. 
	}
	\label{fig:reco}
\end{figure}

%

\else
 
\fi

\section{Discussion}
\label{sec:discussion}
\ifdefined\notFOREPJ
 
We discuss the potential of the track-based multiple scattering tomography in terms of contrast and resolution.
%
%
From Fig.~\ref{fig:reco} the contrast is calculated.
Within a region of air, indicated by the red box, the mean signal of the inverse transform amounts to $\mu_\textrm{air} = \num{0.005(13)}$,
 whereas for the region containing aluminium only the mean signal is $\mu_\textrm{alu} = \num{0.147(18)}$.
At a slice width of $\SI{200}{\um}$, this results in a contrast $C$ of aluminium to air of

\begin{equation}
 C = \frac{\mu_{\textrm{alu}} - \mu_{\textrm{air}}}{\sqrt{\sigma_{\textrm{alu}}^{2} + \sigma_{\textrm{air}}^{2}}} = \num{6.57(19)},
\end{equation}

\noindent
with $\sigma_{\textrm{alu}}$ and $\sigma_{\textrm{air}}$ representing the standard deviation of the signal of the inverse radon transform within the mentioned areas. 


We define the edge resolution $\sigmaedge$ as the width of the transition from one continuous region to another.
This value is extracted by fitting a modified error function
\begin{equation}
 f(z) = k\int_0^z \exp{\left(-\frac12\left(\frac{z'-z_0}{\sigmaedge}\right)^2\right)}\dd z' + b
\end{equation}
to the signal of the inverse radon transform along the $z$-coordinate at three different locations, cf.\ Fig.~\ref{fig:reco}. 
This resembles the integration of a normal distribution with $\sigmaedge$ as its width, a scaling factor $k$, the inflection point $z_0$ and an offset $b$.
To increase the signal-to-noise ratio and therefore the precision of this measurement, the signal is integrated over five voxels in the $y$-direction prior to the fit. 
This does not change the value of $\sigmaedge$, but decreases its uncertainty. 
Hence, for a voxel size of $\SI{80}{\um}$ in the $z$-direction, an edge resolution of $\SI{46.2(27)}{\um}$, $\SI{54.5(30)}{\um}$ and $\SI{52.3(29)}{\um}$
 is extracted for the positions (1), (2) and (3), respectively. 
Smaller voxel sizes are found to result in an even better resolution, but at a constant particle flux the contrast strongly decreases with lower voxel sizes due to higher statistical uncertainties.

Assuming a 6\,mm thick aluminium plate, which corresponds to an $\epssut = 0.067$, a track resolution at the centre of the SUT of $\SI{2.9(1)}{\um}$ is calculated~\cite{gbltool},
 posing a lower limit on the resolvable feature size.
Furthermore, a transversal movement of the traversing particles inside the SUT due to the scattering processes induces a deterioration of the position resolution depending on the thickness of the scatterer. 
The contribution of this effect is calculated to be less than $\SI{3.7}{\um}$ for the simulated sample. 
This implies possible resolutions down to $\SI{5}{\um}$ for this method, although cell sizes of this order come at the cost of the demand of high statistics and therefore,
 in a real experiment, increased measurement time.

Technologically, larger samples could be reconstructed using sensors of larger area.
Samples consisting of materials with shorter radiation length could be dealt with by tuning the particle energy according to the expected energy loss in the sample
 and thereby ensuring a complete transition. 



\else
 
\fi

\section{Conclusion}
\label{sec:conclusion}
\ifdefined\notFOREPJ
 
We demonstrated the feasibility of the reconstruction of centimetre-sized objects probed by high-energy electrons. 
Therefor, the effective angular deflection caused by multiple Coulomb scattering is measured for simulated electron trajectories.
The simulation performed incorporates all necessary physical processes and is a realistic representation of a set-up available at beam lines today. 
The spatial resolution on the SUT is calculated to $\SI{2.9(1)}{\um}$, and allows in principle for the resolution of features on the micrometer scale. 
In order to represent statistics comparable to a real experiment, 280 million electrons were simulated and a voxel size of $200\times80\times\SI{80}{\um^3}$ was chosen,
 leading to resolutions of about $\sigmaedge = \SI{50(2)}{\um}$.
This allows for the discrimination of structures of 0.2\,mm size. 
A contrast of about $C=\num{6.6(2)}$ has been measured comparing regions of aluminium with regions of air.
Possible improvements of contrast and resolution are expected to be obtained by using a dedicated model for the trajectory of the electron,
 including the scattering at the sensor planes of the beam telescope, and a more sophisticated image reconstruction method.
The method is in principle scalable to larger samples by increasing the sensor size, and to shorter radiation lengths by increasing the beam energy,
 allowing this track-based multiple scattering tomography to be used in non-destructive material testing. 


\else
 
\fi


{\small
\section*{Data and materials}
The datasets supporting the conclusions of this article are available from reference \cite{schuetze_data_TBMST}.
The software used is available from the github repositories: \\
 1) \url{https://github.com/eutelescope/eutelescope}, \\
 2) \url{https://github.com/scikit-image/scikit-image}, and \\
 3) \url{https://github.com/simonspa/resolution-simulator}. \\
}


\ifdefined\notFOREPJ
\bibliographystyle{IEEEtran}
\else
\fi
\bibliography{bibtex/refs}

\begin{thebibliography}{10}
\providecommand{\url}[1]{#1}
\csname url@samestyle\endcsname
\providecommand{\newblock}{\relax}
\providecommand{\bibinfo}[2]{#2}
\providecommand{\BIBentrySTDinterwordspacing}{\spaceskip=0pt\relax}
\providecommand{\BIBentryALTinterwordstretchfactor}{4}
\providecommand{\BIBentryALTinterwordspacing}{\spaceskip=\fontdimen2\font plus
\BIBentryALTinterwordstretchfactor\fontdimen3\font minus
  \fontdimen4\font\relax}
\providecommand{\BIBforeignlanguage}[2]{{%
\expandafter\ifx\csname l@#1\endcsname\relax
\typeout{** WARNING: IEEEtran.bst: No hyphenation pattern has been}%
\typeout{** loaded for the language `#1'. Using the pattern for}%
\typeout{** the default language instead.}%
\else
\language=\csname l@#1\endcsname
\fi
#2}}
\providecommand{\BIBdecl}{\relax}
\BIBdecl

\bibitem{Cormack}
\BIBentryALTinterwordspacing
A.~M. Cormack, ``Reconstruction of densities from their projections, with
  applications in radiological physics,'' \emph{Physics in Medicine and
  Biology}, vol.~18, no.~2, p. 195, 1973. [Online]. Available:
  \url{http://stacks.iop.org/0031-9155/18/i=2/a=003}
\BIBentrySTDinterwordspacing

\bibitem{Hounsfield}
\BIBentryALTinterwordspacing
G.~N. Hounsfield, ``Computerized transverse axial scanning (tomography): Part
  1. description of system,'' \emph{The British Journal of Radiology}, vol.~46,
  no. 552, pp. 1016--1022, 1973, pMID: 4757352. [Online]. Available:
  \url{http://dx.doi.org/10.1259/0007-1285-46-552-1016}
\BIBentrySTDinterwordspacing

\bibitem{Ambrose}
\BIBentryALTinterwordspacing
J.~Ambrose, ``Computerized transverse axial scanning (tomography): Part 2.
  clinical application,'' \emph{The British Journal of Radiology}, vol.~46, no.
  552, pp. 1023--1047, 1973, pMID: 4757353. [Online]. Available:
  \url{http://dx.doi.org/10.1259/0007-1285-46-552-1023}
\BIBentrySTDinterwordspacing

\bibitem{HENKE1993181}
\BIBentryALTinterwordspacing
B.~Henke, E.~Gullikson, and J.~Davis, ``X-ray interactions: Photoabsorption,
  scattering, transmission, and reflection at e = 50-30,000 ev, z = 1-92,''
  \emph{Atomic Data and Nuclear Data Tables}, vol.~54, no.~2, pp. 181 -- 342,
  1993. [Online]. Available:
  \url{http://www.sciencedirect.com/science/article/pii/S0092640X83710132}
\BIBentrySTDinterwordspacing

\bibitem{PhysRevD.86.010001}
J.~Behringer et~al. (Particle Data~Group), ``{Review of Particle Physics},''
  \emph{Phys. Rev. D}, vol.~86, p. 010001, Jul 2012.

\bibitem{ref:PDG-2014}
K.~Olive \emph{et~al.}, ``Review of {P}article {P}hysics,'' \emph{Chin. Phys.
  C}, vol.~38, p. 090001, 2014.

\bibitem{nanoCT}
A.~Tkachuk, F.~Duewer, H.~Cui \emph{et~al.}, ``{X-ray computed tomography in
  Zernike phase contrast mode at 8 keV with 50-nm resolution using Cu rotating
  anode X-ray source},'' \emph{{Zeitschrift für Kristallographie - Crystalline
  Materials}}, vol. 222, pp. 650--655, Nov. 2007.

\bibitem{CTcontrastsoft}
K.~A. Faraj \emph{et~al.}, ``{Micro-computed tomographical imaging of soft
  biological materials using contrast techniques},'' \emph{Tissue Eng Part C
  Methods}, vol.~15, pp. 493--499.

\bibitem{Cullen:1989wd}
D.~E. Cullen, M.~H. Chen, J.~H. Hubbell, S.~T. Perkins, E.~F. Plechaty, J.~A.
  Rathkopf, and J.~H. Scofield, ``{Tables and graphs of photon interaction
  cross-sections from 10-eV to 100-GeV derived from the LLNL evaluated photon
  data library (EPDL). Part A: Z = 1 to 50},'' 1981.

\bibitem{PET}
\BIBentryALTinterwordspacing
M.~M. Ter-Pogossian, M.~E. Phelps, E.~J. Hoffman, and N.~A. Mullani, ``A
  positron-emission transaxial tomograph for nuclear imaging (pett),''
  \emph{Radiology}, vol. 114, no.~1, pp. 89--98, 1975, pMID: 1208874. [Online].
  Available: \url{https://doi.org/10.1148/114.1.89}
\BIBentrySTDinterwordspacing

\bibitem{MRI}
P.~C. Lauterbur, ``{Image Formation by Induced Local Interactions: Examples
  Employing Nuclear Magnetic Resonance},'' \emph{Nature}, vol. 242, pp.
  190--191.

\bibitem{OATLEY1966181}
\BIBentryALTinterwordspacing
C.~Oatley, W.~Nixon, and R.~Pease, ``Scanning electron microscopy,''
  \emph{Advances in Electronics and Electron Physics}, vol.~21, pp. 181 -- 247,
  1966. [Online]. Available:
  \url{http://www.sciencedirect.com/science/article/pii/S0065253908610100}
\BIBentrySTDinterwordspacing

\bibitem{TEMsimple}
\BIBentryALTinterwordspacing
A.~V. Crewe, M.~Isaacson, and D.~Johnson, ``A simple scanning electron
  microscope,'' \emph{Review of Scientific Instruments}, vol.~40, no.~2, pp.
  241--246, 1969. [Online]. Available:
  \url{http://dx.doi.org/10.1063/1.1683910}
\BIBentrySTDinterwordspacing

\bibitem{STOLZENBERG2017173}
\BIBentryALTinterwordspacing
U.~Stolzenberg, A.~Frey, B.~Schwenker, P.~Wieduwilt, C.~Marinas, and
  F.~Lütticke, ``Radiation length imaging with high-resolution telescopes,''
  \emph{Nucl. Instr. Meth. Phys. Res. A}, vol. 845, pp. 173 -- 176, 2017,
  proceedings of the Vienna Conference on Instrumentation 2016. [Online].
  Available:
  \url{http://www.sciencedirect.com/science/article/pii/S0168900216306519}
\BIBentrySTDinterwordspacing

\bibitem{Moliere:1948zz}
G.~Moliere, ``{Theorie der Streuung schneller geladener Teilchen II, Mehrfach-
  und Vielfachstreuung},'' \emph{Z. Naturforsch.}, vol.~3a, pp. 78--97, 1948.

\bibitem{PhysRev.89.1256}
\BIBentryALTinterwordspacing
H.~A. Bethe, ``Moli\`ere's theory of multiple scattering,'' \emph{Phys. Rev.},
  vol.~89, pp. 1256--1266, Mar 1953. [Online]. Available:
  \url{https://link.aps.org/doi/10.1103/PhysRev.89.1256}
\BIBentrySTDinterwordspacing

\bibitem{ref:deans2007radon}
\BIBentryALTinterwordspacing
S.~Deans, \emph{The Radon Transform and Some of Its Applications}, ser. Dover
  Books on Mathematics Series.\hskip 1em plus 0.5em minus 0.4em\relax Dover
  Publications, 2007. [Online]. Available:
  \url{https://books.google.de/books?id=xSCc0KGi0u0C}
\BIBentrySTDinterwordspacing

\bibitem{JansenEPJ}
\BIBentryALTinterwordspacing
H.~Jansen \emph{et~al.}, ``{{Performance of the EUDET-type beam telescopes}},''
  \emph{EPJ Techn. Instrum.}, vol.~3, no.~1, p.~7, 2016, {{DESY-16-055,
  arXiv:1603.09669}}. [Online]. Available:
  \url{10.1140/epjti/s40485-016-0033-2}
\BIBentrySTDinterwordspacing

\bibitem{Allpix-github}
M.~Benoit \emph{et~al.}, ``{AllPix},'' \url{https://github.com/ALLPix/allpix},
  accessed: 11.04.2017.

\bibitem{Agostinelli2003250}
\BIBentryALTinterwordspacing
S.~Agostinelli \emph{et~al.}, ``{GEANT4 - a simulation toolkit},''
  \emph{Nuclear Instruments and Methods in Physics Research Section A:
  Accelerators, Spectrometers, Detectors and Associated Equipment}, vol. 506,
  no.~3, pp. 250 -- 303, 2003. [Online]. Available:
  \url{http://www.sciencedirect.com/science/article/pii/S0168900203013688}
\BIBentrySTDinterwordspacing

\bibitem{1610988}
J.~Allison \emph{et~al.}, ``{GEANT4 developments and applications},''
  \emph{IEEE Transactions on Nuclear Science}, vol.~53, no.~1, pp. 270--278,
  Feb 2006.

\bibitem{DESYtb}
R.~Diener, N.~Meyners, N.~Potylitsina-Kube, and M.~Stanitzki, ``{Test Beams at
  DESY},'' \url{http://testbeam.desy.de}, accessed: 26.07.2016.

\bibitem{HuGuo2010480}
\BIBentryALTinterwordspacing
C.~Hu-Guo \emph{et~al.}, ``First reticule size {MAPS} with digital output and
  integrated zero suppression for the {EUDET-JRA1} beam telescope,''
  \emph{Nucl. Instrum. Methods Phys. Rev. A}, vol. 623, no.~1, pp. 480 -- 482,
  2010, 1st International Conference on Technology and Instrumentation in
  Particle Physics. [Online]. Available: \url{10.1016/j.nima.2010.03.043}
\BIBentrySTDinterwordspacing

\bibitem{1748-0221-11-12-C12031}
\BIBentryALTinterwordspacing
H.~Jansen, ``{Resolution studies with the DATURA beam telescope},'' \emph{J.
  Inst.}, vol.~11, no.~12, p. C12031, 2016. [Online]. Available:
  \url{http://stacks.iop.org/1748-0221/11/i=12/a=C12031}
\BIBentrySTDinterwordspacing

\bibitem{ref:scatteringhighland}
V.~Highland, ``Some practical remarks on multiple scattering,'' \emph{Nucl.
  Instrum. Methods Phys. Rev. A}, vol. 129, no.~2, pp. 497--499, 1975.

\bibitem{Berger}
\BIBentryALTinterwordspacing
N.~Berger, A.~Buniatyan, P.~Eckert, F.~Förster, R.~Gredig, O.~Kovalenko,
  M.~Kiehn, R.~Philipp, A.~Schöning, and D.~Wiedner, ``Multiple coulomb
  scattering in thin silicon,'' \emph{Journal of Instrumentation}, vol.~9,
  no.~07, p. P07007, 2014. [Online]. Available:
  \url{http://stacks.iop.org/1748-0221/9/i=07/a=P07007}
\BIBentrySTDinterwordspacing

\bibitem{schuetze_data_TBMST}
P.~Sch\"utze and H.~Jansen, ``{Dataset for the `Feasibility of track-based
  multiple scattering tomography'},'' \url{10.5281/zenodo.852408}, Aug. 2017.

\bibitem{ref:RadonOrig}
J.~Radon, ``{\"U}ber die {B}estimmung von {F}unktionen durch ihre
  {I}ntegralwerte l\"angs gewisser {M}annigfaltigkeiten,'' \emph{Akad. Wiss.},
  vol.~69, pp. 262--277, 1917.

\bibitem{ref:scikitWebpage}
{scikit-image development team}, ``{scikit-image},''
  \url{http://scikit-image.org/}, accessed: 24.06.2017.

\bibitem{ref:scikitArticle}
\BIBentryALTinterwordspacing
S.~van~der Walt, J.~L. {S}ch\"onberger, J.~{Nunez-Iglesias}, F.~{B}oulogne,
  J.~D. {W}arner, N.~{Y}ager, E.~{G}ouillart, T.~{Y}u, and the scikit-image
  contributors, ``{scikit-image: image processing in Python},'' \emph{PeerJ},
  vol.~2, p. e453, 6 2014. [Online]. Available:
  \url{http://dx.doi.org/10.7717/peerj.453}
\BIBentrySTDinterwordspacing

\bibitem{gbltool}
\BIBentryALTinterwordspacing
S.~Spannagel and H.~Jansen, ``{GBL Track Resolution Calculator},'' accessed:
  03.03.2016. [Online]. Available:
  \url{https://github.com/simonspa/resolution-simulator}
\BIBentrySTDinterwordspacing

\end{thebibliography}

\end{document}